# COVID-19 in South Asia: Real-time monitoring of reproduction and case fatality rate

Fakhar Mustafa[a,e], Rehan Ahmed Khan Sherwani[a], Syed Salman Saqlain[b], Muhammad Asad Meraj[c], Haseeb ur Rehman[d], Rida Ayyaz[e]

## Abstract:

As the ravages caused by COVID-19 pandemic are becoming inevitable with every moment, monitoring and understanding of transmission and fatality rate has become even more paramount for containing its spread. The key purpose of this analysis is to report the real-time effective reproduction rate ($R_t$) and case fatality rates (CFR) of COVID-19 in South Asia region. Data for this study are extracted from JHU CSSE COVID-19 Data source up to July 31, 2020. $R_t$ is estimated using exponential growth and time-dependent methods. 'R0' package in R-language is employed to estimate $R_t$ by fitting the existing epidemic curve. Case fatality rate is estimated by using naïve and Kaplan-Meier methods. Owing to exponential increase in cases of COVID-19, the pandemic will ensue in India, Maldives and in Nepal as $R_t$ was estimated greater than 1 for these countries. Although case fatality rates are found lesser as compared to other highly affected regions in the world, strict monitoring of deaths for better health facilities and care of patients is emphasized. More regional level cooperation and efforts are the need of time to minimize the detrimental effects of the virus.

## Introduction:

Coronavirus disease (COVID-19) is a severe acute respiratory syndrome which has affected approximately 18 million people and caused more than 0.68 million deaths and has spread over in 213 countries and territories as of July 31, 2020. COVID-19 infection is fundamentally transmitted through respiratory droplets and contact between individuals from coughing or sneezing as of recent evidence.[1] After exposure to the virus, it could take 2 to 14 days for appearance of symptoms which are more often dry cough, fever, tiredness, difficulty in breathing and muscle aches. Albeit the vast majority with COVID-19 have mild to moderate symptoms, the disease can make extreme clinical inconveniences and cause death in some people. People with existing medical chronic conditions and old age people face more serious danger of getting seriously ill with COVID-19.

[a] *College of Statistics and Actuarial Sciences, University of the Punjab, Pakistan.*
[b] *Department of Orthopaedic Surgery, Lahore General Hospital, Pakistan*
[c] *University of the Education, Multan Campus, Pakistan.*
[d] *Social Welfare and Bait-ul-Maal Department, Government of Punjab, Pakistan.*
[e] *COMSATS University Islamabad, Sahiwal Campus, Pakistan.*

South Asia is one of world's densely populated and least developed region, and is highly prone to breakout of any major infectious disease majorly due to high population rate and compromised medical facilties.[2] COVID-19 outbreak in South Asia is a major public health concern. Local transmission of the virus has been confirmed in all the countries of South Asia. South Asia region reported its first confirmed case on January 24, 2020 in Nepal. As of July 31, 2020, more than 2.33 million confirmed cases of COVID-19 with 47,872 cumulative deaths have been reported in the region. More than 1.57 million people have recovered and there are 0.71 million COVID-19 cases active in the region.

In term of confirmed cases, as of July 31, 2020, India has reported more than 1.75 million confirmed cases of COVID-19, followed by Pakistan and Bangladesh. In term of deaths, India has reported 37,403 deaths which comprises 78.1% of total deaths in the region, followed by Pakistan and Bangladesh which has reported 5,970 and 3,132 deaths, respectively. Details for confirmed cases, deaths, recoveries and active cases for all countries in South Asia region is provided in **Table 1**.

Countries in the region have embraced certain measures to contain and forestall the transmission of COVID-19, since the reporting of first case. Afghanistan temporarily closed its land border with other countries and imposed restrictions on public gatherings. Maldives declared a public health emergency, placed temporary travel restriction and also imposed lockdown in different parts of country. Bangladesh imposed nationwide lock down till June, 2020 and then reimpose area wise lockdowns in the country. Bhutan sealed off land borders and also issue travel advisory. India observed a voluntary public curfew on March 22, 2020 which was followed by 21 days nationwide lockdown on March 24, 2020. The lockdown was then extended till June 1, 2020 with substantial relaxations in the month of May, 2020. Nepal observed the strategy of countrywide lockdown and also imposed travel restrictions. In Pakistan, Gilgit Baltistan government was the first to impose complete lockdown on March 22, 2020 followed by Khyber Pakhtunkhwa province on March 23,2020. On March 24, 2020 all remaining provinces and territories imposed complete lockdown which was then extended up to May 11,2020. Since the mid of June, 2020, the government is following a smart lockdown strategy to reduce the spread of COVID-19. Sri Lanka prohibited all foreign arrivals at least until August and also enforced restrictions on mass gatherings. On March 15, 2020, governments in South Asia coordinated a response to the COVID-19 pandemic through South Asia Association for Regional Cooperation (SAARC) organization and set up SAARC COVID-19 Emergency Fund.

In spite of local and regional efforts number of COVID-19 confirmed cases and deaths are increasing exponentially in the region. The implementation of time-sensitive preventive and control strategies depending upon continuously monitoring the transmission and fatality rate of virus in is need of time especially in least developed region like South Asia. Present work on COVID-19 transmission and fatality rate has focused on China, Europe and other regions. Liu et al. (2020) estimated highest reproduction rate as compared to rate estimated by WHO in China.[3] Zhang et al. (2020) performed a data-driven analysis to estimate the reproductive number of novel coronavirus and its probable outbreak size on the Diamond Princess cruise ship.[4] Jing et al. (2020) provides significant findings in understanding the dynamics of COVID-19 early outbreak in four most affected countries of Europe.[5] Khosravi et al. (2020) observed a significant decrease in reproduction rate of COVID-19 due to success of interventions imposed by health system and also predicted the expected number of new cases in Shahroud in Northeastern Iran.[6]

No attention has been paid to alarming condition of South Asia region. In this paper we aim to calculate real time effective reproduction and case fatality rate given the continuous rise in cases of COVID-19 in South Asia. This study will help understanding the dynamics of transmission and mortality of the virus and will also assist governments in the region to implement control measures and prioritize preventions in the region. Reproduction rate ($R_t$) measures the effective reproduction number of infectious disease when immunity steps are taken. On the other hand, the basic reproduction number ($R_0$) is calculated where there has been no immunity from prior infection or vaccinations, or any direct interference in the spread of the disease. It is appropriate to use $R_t$ to monitor real time transmission of disease amid public health interventions. Mortality rate can be measured by using the case fatality rate (CFR).

**Data Source:**

Data were obtained from routine COVID-19 laboratory-confirmed cases, made publicly accessible by the Center for Systems Science and Engineering (CSSE) at Johns Hopkins University (https://github.com/CSSEGISandData/COVID-19). Statistics up to July 31, 2020 have been used for this study.

**Table 1: COVID-19 Report for countries in South Asia region as of July 31, 2020**

| Country | Confirmed Cases | Reported Deaths | Patients Recovered | Active Cases |
|---|---|---|---|---|
| Afghanistan | 36,710 | 1,283 | 25,509 | 9,918 |
| Bangladesh | 239,860 | 3,132 | 136,253 | 100,475 |
| Bhutan | 101 | 0 | 89 | 12 |
| India | 1,751,919 | 37,403 | 1,146,879 | 567,637 |
| Maldives | 3,949 | 17 | 2,613 | 1319 |
| Nepal | 20,086 | 56 | 14,492 | 5,538 |
| Pakistan | 279,146 | 5,970 | 248,027 | 25,149 |
| Sri Lanka | 2,815 | 11 | 2,439 | 365 |
| **Total** | **2,334,586** | **47,872** | **1,576,301** | **710,413** |

## Methods and Material:

**Estimation of the Reproduction Number:**

**Reproduction Number $R_t$:**

In such a population which is entirely susceptible of an infectious disease, the average number of secondary cases of an infectious disease that one case would produce is denoted by $R_0$.[7] Standard compartmental models delineate that how $R_0$ is identified with the average age of infection, vaccination limits for elimination and equilibrium solutions. In any case, a large number of the essential formulae for $R_0$ separate when we believe transmission of infection to be a stochastic procedure including discrete individuals.[8]

The real-time reproduction number $R_t$, characterized as the number of secondary cases that would produce over the span of the outbreak, is helpful to screen the transmissibility of COVID-19 over the time.[9]

The exponential growth rate (EG) method and the time-dependent (TD) method were used to estimate $R_t$ values.

**Exponential Growth Method (EG):**

In EG method, based on the assumption of generation time through gamma distribution, the reproduction rate is estimated by converting the exponential growth rate and by changing the Poisson regression model to the exponential growth point of the outbreak of the disease.[3, 9]

**Time Dependent Method (TD):**

The TD method measures reproduction numbers by means of an average of all propagation networks consistent with observations.[9] This approach incorporates a Bayesian statistical paradigm, in which continuous calculation often takes into account unrecorded cases prior to end of the epidemic. Additionally, this method provides good epidemic curve fitting.[5]

However, the methods referred vide supra, require generation time (GT) that is determined by taking the time difference between primary and secondary cases, which cannot be easily obtained. In this study, the GT was assumed to be equal to the incubation period, which was estimated to be 5.8 days (standard deviation (SD) 2.6 days).[10] An estimate of 4 days (SD 2.4 days) was also used in the sensitivity analysis.[11]

**Estimation of the Case Fatality Rate:**

**Case Fatality Rate (CFR):**

The case fatality rate is the percentage of deaths of patients from any disease or injury in a certain period of time. This likelihood provides knowledge about the disease's prognosis and the effectiveness of therapeutic mediation.[12]

Naive estimates of CFR from the reported numbers of confirmed cases and deaths are difficult to interpret due to the possible under ascertainment of mild cases and the right-censoring of cases with respect to the time delay from illness onset to death.[13] One method of overcoming this is to compute CFR by utilizing the Kaplan-Meier method for use with two outcomes that are death and recovery, over the span of an epidemic.[14] In real-time CFR is estimated using two different methodologies, the naïve method (*nCFR*) and the Kaplan-Meier method for calculating CFR (*kCFR*), where

$$nCFR = \frac{Cumulative\ Deaths}{Cumulative\ Cases}$$

and

$$kCFR = \frac{Cumulative\ Deaths}{(Cumulative\ Recoveries + Cumulative\ Deaths)}$$

**R0 Package:**

In order to estimate the reproduction rate for COVID-19 in South Asia, the R0 package an R-language coded statistical package was used.[15] This software package allows a standardized and extensible approach to the estimation of the reproduction number and the generation interval distribution from epidemic curves. All the results in this study were computed through R-language.[16]

## Results:

Reproduction rate of COVID-19 for each country in the South Asia region has been calculated from the onset of first confirmed cases till July 31, 2020. Different starting periods for each country have been selected considering the increment in confirmed cases on a smooth scale.

Initially, all possible combinations of starting and ending dates were estimated under exponential growth method (EG) (**Table 2**). For Afghanistan, the period beginning at March 08 and ending at March 31, provided highest $R_t$ for COVID-19 that was estimated to be 2.2 (95% CI: 1.93-2.60) under GT1 and 1.82 (95% CI: 1.63-2.06) under GT2. Lowest reproduction rate was observed in the period between July 1 and July31 which is estimated as 0.74 (95% CI 0.73-0.76) under GT1 and 0.81 (95% CI 0.80-0.82) under GT2. For Bangladesh, highest reproduction rate was observed for the period between April 1 and April 3, corresponding $R_t$ value 1.79 (95% CI: 1.76-1.83) under GT1 and 1.54 (95% CI: 1.52-1.56) under GT2. Lowest reproduction rate was observed in the period between July 1 and July 31 with estimated $R_t$ of 0.94 (95% CI: 0.94-0.95) under GT1 and 0.96 (95% CI: 0.95-0.96) under GT2. For Bhutan, highest reproduction rate was found between the period of May 1 and May 31, which was estimated to be 1.72 (95% CI: 1.35-2.23) under GT1 and 1.49 (95% CI: 1.24-1.86) under GT2. Lowest reproduction rate was observed in the period of June 1 and June 30, with estimated $R_t$ of 0.81 (95% CI: 0.62-1.04) under GT1 and 0.86 (95% CI: 0.71-1.03) under GT2. India which is the most affected country in South Asian region reported highest reproduction rate for the period between March 02 to March 31 which was 2.2 (95% CI: 2.14-2.35) under GT1 and 1.83 (95% CI: 1.76-1.89) under GT2. Lowest reproduction rate was observed in between the period of June 1 and June 30, with estimated $R_t$ of 1.20 (95% CI: 1.20-1.21) under GT1 and 1.14 (95% CI: 1.14-1.15) under GT2. For Maldives, lowest estimated $R_t$ was observed for the period starting on March 03 and ending at March 31, which was 0.59 (95% CI: 0.34-0.94) under GT1 and 0.69 (95% CI: 0.48-0.96) under GT2. Meanwhile, highest reproduction rate was observed 4.16 (95% CI: 3.7-4.7) under GT1 and 3.02 (95% CI: 2.74-3.34) under GT2 in between the time period of April 1 and April 30. For Nepal, highest reproduction rate was observed in the period of May 1 and May 31, with corresponding $R_t$ of 2.12 (95% CI: 2.03-2.21) under GT1 and 1.75 (95% CI: 1.69-1.81) under GT2. Lowest reproduction rate was observed in the period starting from February 23 and ending at March 31, with estimated $R_t$ of 0.24 (95% CI: 0 -2.99) under GT1 and 0.39 (95% CI: 0.03-2.30) under GT2. For Pakistan, the period starting on February 25 and ending on March 31, yielded highest $R_t$ that was estimated 2.03 (95% CI: 1.96-2.10) under GT1 and 1.69 (95% CI: 1.65-1.74) under GT2. Lowest reproduction rate was observed in the period between July 1 and July31, with estimated $R_t$ of 0.71 (95% CI: 0.70-0.71) under GT1 and 0.79 (95% CI: 0.78-0.79) under GT2. For Sri Lanka, highest reproduction rate was observed in the period between April 1 and April 30, with corresponding $R_t$ of 1.76 (95% CI: 1.65-1.89) under GT1 and 1.52 (95% CI: 1.44-1.60) under GT2. Lowest reproduction rate was observed from June 1 and June 30, with estimated $R_t$ of 0.68 (95% CI: 0.63-0.74) under GT1 and 0.77 (95% CI: 0.72-0.81) under GT2.

**Table 2: Estimated $R_t$ values through Exponential Growth Method for GT1 and GT2 time intervals**

| | | | EG | | EG | |
|---|---|---|---|---|---|---|
| Country | Start Date | End Date | **GT1 = 5.6 days** | | **GT2 = 4 days** | |
| | | | $R_t$ | 95% C.I | $R_t$ | 95% CI |
| Afghanistan | 08-03-2020 | 31-03-2020 | 2.24 | 1.93 - 2.6 | 1.82 | 1.63 - 2.06 |
| | 01-04-2020 | 30-04-2020 | 1.46 | 1.42 - 1.51 | 1.32 | 1.29 - 1.35 |
| | 01-05-2020 | 31-05-2020 | 1.35 | 1.34 - 1.37 | 1.25 | 1.23 - 1.26 |
| | 01-06-2020 | 30-06-2020 | 0.79 | 0.78 - 0.80 | 0.85 | 0.84 - 0.86 |
| | 01-07-2020 | 31-07-2020 | 0.73 | 0.71 - 0.75 | 0.80 | 0.78 - 0.82 |
| Bangladesh | 08-03-2020 | 31-03-2020 | 1.26 | 0.98 - 1.61 | 1.18 | 0.99 - 1.42 |
| | 01-04-2020 | 30-04-2020 | 1.79 | 1.76 - 1.83 | 1.54 | 1.52 - 1.56 |
| | 01-05-2020 | 31-05-2020 | 1.32 | 1.31 - 1.33 | 1.22 | 1.21 - 1.23 |
| | 01-06-2020 | 30-06-2020 | 1.09 | 1.08 - 1.09 | 1.06 | 1.06 - 1.06 |
| | 01-07-2020 | 31-07-2020 | 0.93 | 0.92 - 0.94 | 0.95 | 0.94 - 0.96 |
| Bhutan | 06-03-2020 | 31-03-2020 | 1.24 | 0.52 - 2.86 | 1.17 | 0.64 - 2.22 |
| | 01-04-2020 | 30-04-2020 | 0.98 | 0.37 - 2.22 | 0.99 | 0.51 - 1.81 |
| | 01-05-2020 | 31-05-2020 | 1.72 | 1.35 - 2.23 | 1.49 | 1.24 - 1.82 |
| | 01-06-2020 | 30-06-2020 | 0.81 | 0.62 - 1.04 | 0.86 | 0.71 - 1.03 |
| | 01-07-2020 | 31-07-2020 | 1.31 | 0.81 - 2.10 | 1.31 | 0.81 - 2.10 |
| India | 02-03-2020 | 31-03-2020 | 2.24 | 2.14 - 2.35 | 1.83 | 1.76 - 1.89 |
| | 01-04-2020 | 30-04-2020 | 1.33 | 1.32 - 1.34 | 1.23 | 1.22 - 1.23 |
| | 01-05-2020 | 31-05-2020 | 1.27 | 1.27 - 1.28 | 1.19 | 1.19 - 1.20 |
| | 01-06-2020 | 30-06-2020 | 1.20 | 1.20 - 1.21 | 1.14 | 1.14 - 1.15 |
| | 01-07-2020 | 31-07-2020 | 1.26 | 1.25 - 1.26 | 1.18 | 1.18 - 1.18 |
| Maldives | 08-03-2020 | 31-03-2020 | 0.59 | 0.34 - 0.94 | 0.69 | 0.48 - 0.96 |
| | 01-04-2020 | 30-04-2020 | 4.16 | 3.70 - 4.70 | 3.02 | 2.74 - 3.34 |
| | 01-05-2020 | 31-05-2020 | 1.17 | 1.13 - 1.22 | 1.12 | 1.09 - 1.15 |
| | 01-06-2020 | 30-06-2020 | 0.97 | 0.91 - 1.03 | 0.98 | 0.94 - 1.02 |
| | 01-07-2020 | 31-07-2020 | 1.17 | 1.10 - 1.25 | 1.25 | 1.07 - 1.18 |
| Nepal | 23-03-2020 | 31-03-2020 | 0.24 | 0.00 - 2.99 | 0.39 | 0.03 - 2.30 |
| | 01-04-2020 | 30-04-2020 | 1.30 | 1.07 - 1.59 | 1.21 | 1.05 - 1.40 |
| | 01-05-2020 | 31-05-2020 | 2.12 | 2.03 - 2.21 | 1.75 | 1.69 - 1.81 |
| | 01-06-2020 | 30-06-2020 | 1.17 | 1.15 - 1.18 | 1.12 | 1.11 - 1.13 |
| | 01-07-2020 | 31-07-2020 | 0.59 | 0.57 - 0.61 | 0.69 | 0.67 - 0.70 |
| Pakistan | 25-02-2020 | 31-03-2020 | 2.03 | 1.96 - 2.10 | 1.69 | 1.65 - 1.74 |
| | 01-04-2020 | 30-04-2020 | 1.41 | 1.39 - 1.43 | 1.28 | 1.27 - 1.29 |
| | 01-05-2020 | 31-05-2020 | 1.18 | 1.17 - 1.19 | 1.12 | 1.12 - 1.13 |
| | 01-06-2020 | 30-06-2020 | 0.94 | 0.93 - 0.94 | 0.95 | 0.95 - 0.96 |
| | 01-07-2020 | 31-07-2020 | 0.71 | 0.70 - 0.72 | 0.79 | 0.78 - 0.79 |
| Sri Lanka | 11-03-2020 | 31-03-2020 | 1.15 | 0.97 - 1.36 | 1.11 | 0.98 - 1.25 |
| | 01-04-2020 | 30-04-2020 | 1.76 | 1.65 - 1.89 | 1.52 | 1.44 - 1.60 |
| | 01-05-2020 | 31-05-2020 | 1.40 | 1.34 - 1.46 | 1.27 | 1.23 - 1.32 |
| | 01-06-2020 | 30-06-2020 | 0.68 | 0.63 - 0.74 | 0.77 | 0.72 - 0.81 |
| | 01-07-2020 | 31-07-2020 | 0.89 | 0.83 - 0.96 | 0.92 | 0.87 - 0.97 |

Over time, comparative patterns were observed in $R_t$ with both generated time intervals GT1 and GT2 in TD method (**Figure 1**). For Afghanistan, from March 08 to July 31, $R_t$ was in the range of 0.5–4.0. The $R_t$ as of July 31 was 0.34 (95% CI: 0–1.42) under GT1 and 0.34 (95% CI: 0–2.20) under GT2. For Bangladesh, $R_t$ values were in range of 0.5 and 7.5, and on July 31 it was 0.80 (95% CI: 0.51-1.16) under GT1 and 0.77 (95% CI 0.31–1.37) under GT2. For Bhutan, $R_t$ values were not estimated smoothly due to high number of zero cases for series of many days. However, from March 06 and onwards, $R_t$ values were in the range 0.3-4.4 in Bhutan. For India, from March 02 till July 31, $R_t$ values decreased from 13 to 1.08 and then again it started increasing. Corresponding $R_t$ was 1.12 (95% CI: 1.04-1.20) under GT1 and 1.10 (95% CI: 0.96-1.24) under GT2. For Maldives, $R_t$ values were in range of 0.36-10 between March 08 and July 24. As of July 31, corresponding $R_t$ values was 2.06 (95% CI: 0-6.41) under GT1 and 0.88 (95% CI: 0–3.85) under GT2. For Nepal, $R_t$ values were in the range of 0.6–4. The corresponding $R_t$ value was 1.76 (95% CI: 0.42–3.39) under GT1 and 1.40 (95% CI: 0–3.92) under GT2. For Pakistan, $R_t$ values were in range of 0.6-9.8 between February 25 and July 31. As of July 31, corresponding $R_t$ values were 0.49 (95% CI: 0.11-1.01) under GT1 and 0.52 (95% CI: 0-1.39) under GT2. For Sri Lanka, corresponding $R_t$ was 0.04 (95% CI: 0-0.09) under GT1 method and 0.003 (95% CI: 0-0.01) under GT2 method as of July 31. Corresponding $R_t$ values are zero for all zero cases reported on different dates in **Figure 1**.

In Afghanistan, first death due to COVID-19 was reported on March 22, 2020. As of July 31, nCFR was 0.034 and *kCFR* was 0.047. In Bangladesh, first death was reported on March 18. Highest *kCFR* was observed in the month of April. As of July 31, corresponding *nCFR* and *kCFR* was 0.013 and 0.022 respectively. No death had been reported in Bhutan till July 31. India reported highest number of deaths due to complications of COVID-19 in the region. Corresponding *nCFR* and *kCFR* was 0.021 and 0.031. On April 29, Maldives confirmed first death due to COVID-19. Corresponding *nCFR* and *kCFR* was 0.004 and 0.006. As of July 31, Nepal yielded *nCFR* and *kCFR* at 0.002 and 0.003. On May 18 first death was reported in the country. Pakistan reported second highest number of deaths due to the virus in the region. As of July 31, respective *nCFR* and *kCFR* was 0.021 and 0.023. Sri Lanka, reported its first death due to COVID-19 on March 28. As of July 31, corresponding *nCFR* and *kCFR* were 0.003 and 0.004. The value of the CFR relied on the form of calculation: the *kCFR* provide higher estimation (**Figure 2**).

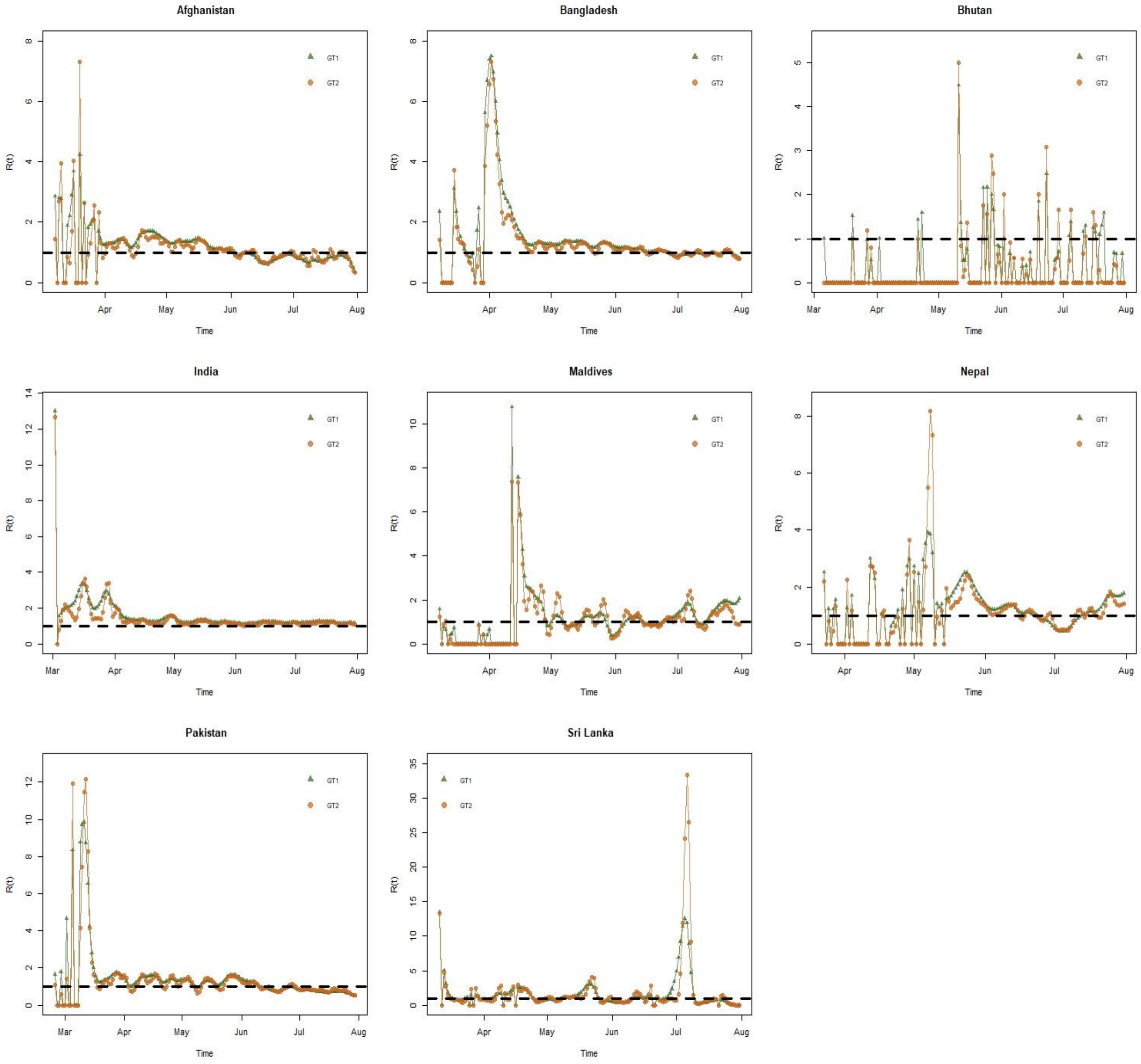

**Figure 1: Time Dependent $R_t$ for countries in South Asia region**

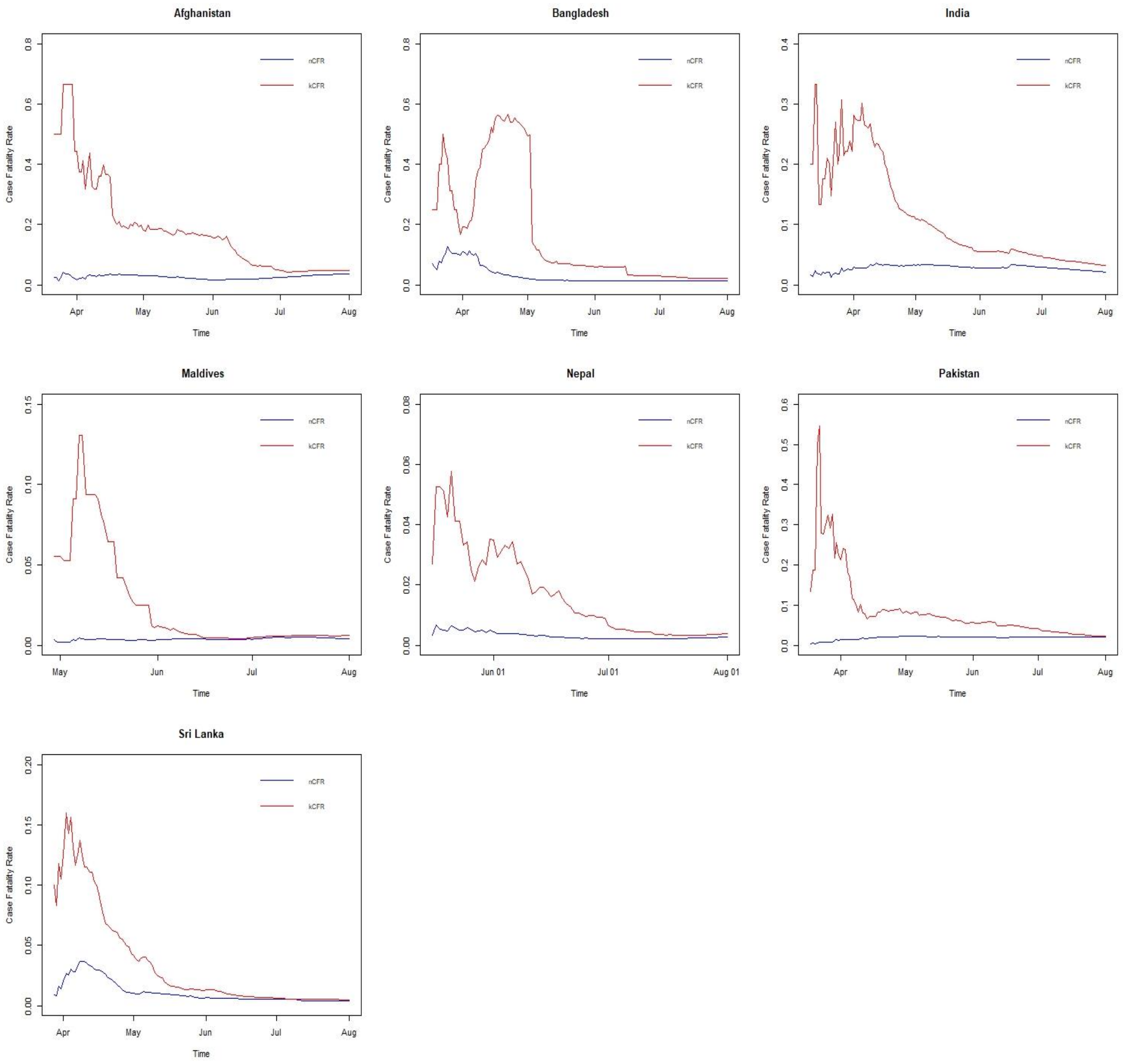

**Figure 2: nCFR and kCFR values for countries in South Asia region**

## Discussion:

Home to one-fifth of the world’s population, South Asia is facing severe health challenges on various fronts. South Asia is second to Africa having a high rate of infant mortality, less expectancy of life, malnutrition, TB and HIV/AIDS cases.[17] One of the most densely populated regions in the world, it also has poor healthcare infrastructure and poor sanitation system. The COVID-19 has compromised South Asia's social and economic conditions very severely leading to crippling of the society. The region will likely to experience its worst economic performance in the last 40 years, with temporary economic contractions in all eight countries of the region.[18] Despite of all the efforts taken to contain the spread of the deadly virus number of cases in the region are increasing exponentially. It is time for all the countries in the South Asia to ponder over on their priorities, resources, capabilities and capacities to come out of most challenging situation of this century. The solution lies in substantial coordination and cooperation.

In order to understand the dynamics of COVID-19, this study gives significant findings on the transmission and fatality rate of COVID-19 in South Asia. As of July 31, in India, Maldives and Nepal reproduction rates were found higher than 1 ($R_t > 1$) depending upon estimation techniques, showing that the outbreak of COVID-19 will proceed into future. In efforts to slow down the spread of virus, certain control measures and progressively severe counteractions are prescribed in these countries. Smart lockdown strategy adopted by Pakistan at different hotspots is producing encouraging outcomes and number of COVID-19 cases are declining in the country.[19] In order to explore the effects of smart lockdown strategy, continued monitoring of new cases of COVID-19 reported in the Pakistan is required. Dynamic surveillance of new cases in the region and in other countries could assist with bettering and comprehending the impacts of various strategies embraced to contain the transmission of virus.

A total of 47,872 deaths were reported in the region till July 31,2020. Corresponding *nCFR* and *kCFR* for whole region was 0.020 and 0.029, respectively. Fatality rate of COVID-19 in the region was found to be lesser as compared to other regions of the world especially highly affected region of Europe (*nCFR*=0.073 and *kCFR*=0.108), South America (*nCFR*=0.036 and *kCFR*=0.051) and North America (*nCFR*=0.041 and *kCFR*=0.076). The lower fatality rate could be attributed to the ongoing nature of the outbreak, distinctive age appropriations, humid climate pattern and diverse treatment techniques adopted in the region. Continued monitoring of deaths caused by COVID-19 in the region is emphasized for better care of patients. In spite of implementation of different strategies, significant increase of confirmed cases has been observed in India and Bangladesh. Real-time tracking of deaths due to newly reported cases of COVID-19 is relevant in those countries of the region in which cases are still increasing.

Significant decrease in reproduction and fatality rate of COVID-19 has been observed over the course of time in South Asia as the result of different preventive strategies and control measures implemented by governments in the region. Mitigations and expansion of case findings should be continued. We strongly suggest implementation of all preventive measures which are suggested by World Health Organization (WHO) to contain the virus especially during upcoming religious and cultural events. Stringent monitoring of close contacts is constantly required during this period. There is a dire need of regional cooperation to combat the adverse effects of COVID-19. In March, through SAARC platform, countries coordinated a response and set up SAARC COVID-19 Emergency Fund to minimize the detrimental effects of the virus but more regional level cooperation and efforts are required. Through active collaboration, mutual assistance, sharing strategies for containing the virus, joint research programs, and routine based active surveillance through SAARC platform, countries in South Asia can accelerate regional efforts like European Center for Disease Prevention and Control (ECDC) to contain socio-economic impact of COVID-19.

.